\documentclass[final,1p,times]{elsarticle}

\usepackage{amssymb}
\usepackage[raggedrightboxes]{ragged2e}

\usepackage{amsmath}
\usepackage{wasysym}

\usepackage[starfontserif]{starfont}
\DeclareSymbolFont{starfontsym}{OT1}{sts}{m}{n}
\DeclareMathSymbol{\mathTerra}{\mathord}{starfontsym}{76}

\usepackage [english]{babel}
\usepackage [autostyle, english = american]{csquotes}
\MakeOuterQuote{"}

\usepackage{color}
\newcommand{\rev}[1]{#1}
\newcommand{\revv}[1]{#1}
\newcommand{\revvv}[1]{#1}
\newcommand{\revminor}[1]{#1}

\journal{Icarus}

\begin{document}

\begin{frontmatter}

\title{Limit cycles and the climate history of Mars}

\author[inst1]{Jacob Haqq-Misra\corref{cor1}}
\ead{jacob@bmsis.org}
\cortext[cor1]{Corresponding author}

\affiliation[inst1]{organization={Blue Marble Space Institute of Science},Department and Organization
            addressline={600 1st Avenue, 1st Floor}, 
            city={Seattle},
            state={WA},
            postcode={98104}, 
            country={USA}}

\begin{abstract}
Evidence for fluvial features and standing liquid water indicate that Mars was a warmer and wetter place in its past; however, climate models have historically been unable to produce conditions to yield a warm early Mars under the faint young sun. Some models invoke thick greenhouse atmospheres to produce continuously warm conditions, but others have argued that available geologic evidence is more consistent with short-duration and transient warming events on an otherwise cold Mars. One possibility of harmonizing these perspectives is that early Mars experienced climate limit cycles that caused the climate to oscillate between short periods of warmth and prolonged periods of glaciation, due to modulation of greenhouse warming by the carbonate-silicate cycle. This study suggests that episodic limit cycling during the Noachian and Hesperian periods provides a hypothetical explanation for the timing and formation of fluvial features on Mars. A schematic time-forward trajectory of the full history of Mars is calculated using an energy balance climate model, which includes an active carbonate-silicate cycle, instellation changes due to the sun's main sequence evolution, variations in the obliquity of Mars, and supplemental warming from additional greenhouse gases beyond carbon dioxide alone. These calculations demonstrate the viability of a climate history for Mars involving episodic limit cycling to enable the formation of the valley networks at 4.1-3.5\,Ga and delta features at 3.3-3.0\,Ga, interspersed with cold stable climates and ending in the late Amazonian in a carbon dioxide condensation regime. This schematic climate trajectory provides a plausible narrative that remains consistent with available geologic data, and further exploration of warming mechanisms for the climate of Mars should consider the possibility of episodic transient events driven by carbonate-silicate limit cycling.
\end{abstract}



\end{frontmatter}

\newpage


\section{Introduction}
\label{sec:introduction}

Evidence that liquid water once flowed on the surface of Mars continues to build, with missions such as the Mars Science Laboratory and the Mars 2020 Perseverance Rover following the fluvial clues first noticed by earlier missions like Mariner 9 and Viking. This geomorphological evidence includes the presence of features such as valley networks and meandering channels \citep[e.g.,][]{cabrol2001composition,fassett2008timing,hynek2010updated,wordsworth2016climate,kite2019geologic}, which appear to be driven by precipitation in an active hydrological cycle \citep[e.g.,][]{barnhart2009long,hynek2010updated}. Crater lakes have also been observed: a striking example is the Mars Science Laboratory discovery of in situ fluvial erosion in Gale Crater, which suggests that the crater itself was filled with liquid water for up to $\sim$10 million years \citep{williams2013martian,grotzinger2015deposition}. Other arguments have suggested that Mars once hosted a northern ocean, with possible evidence of an ancient shoreline \citep{head1999possible,di2010ancient}. The widespread presence of phyllosillicates has also been interpreted as evidence of surface liquid on early Mars, as the formation of phyllosilicates (and other observed aqueous minerals) requires liquid water with near-neutral pH conditions \citep[e.g.,][]{murchie2009synthesis,fairen2011cold,carter2010detection,carter2013hydrous,ehlmann2014mineralogy,wordsworth2016climate}. Mars, at some point in its past, appears to have been a warmer and wetter place.

Most of these fluvial features formed during the late Noachian to early Hesperian periods, about 3.5 to 3.8 Ga, based \rev{on} ages determined from crater counting \citep{fassett2008timing}. The Noachian period ($\sim$3.5 to 4.1 Ga) shows the strongest evidence of standing liquid water, including a northern ocean, while the Hesperian period ($\sim$3.0 to 3.5 Ga) shows some evidence of catastrophic flooding as well as extreme volcanism \citep{wordsworth2016climate,kite2019geologic,kite2019persistence}. The Amazonian period includes Mars today, which is characterized by arid surface conditions and limited weathering, although the early Amazonian may have still shown some activity in its hydrosphere \citep{liu2022zhurong}. This evidence all suggests that the climate of Mars has experienced large-scale and drastic changes, evolving from a habitable world in the early Hesperian into a dry, inhospitable place in the late Amazonian \citep{kite2022high}. 

An analysis by \citet{kite2019geologic} provided quantitative constraints on the presence and duration of an active hydrological cycle, and accompanied formation of rivers or oceans, based upon available geologic observations on Mars. \citet{kite2019geologic} argued that crosscutting relationships among fluvial features as well as crater counts indicate ``high confidence'' that at least two distinct river-forming periods occurred in the history of Mars from the Noachian to early Amazonian. \citet{kite2019geologic} estimated that the total time spanned by river-forming climates on Mars is greater than 10$^{8}$\,yr, with the maximum surface extent of the ocean greater than 10$^{6}$\,km$^{2}$. Likewise, a subsequent analysis of global paleochannel data by \citet{kite2019persistence} concluded that rivers on Mars were wider than those on Earth today, which suggests periods of intense and long-lived runoff on ancient Mars that persisted until about 3 Ga ago. 

The problem of explaining a wet, habitable climate on early Mars has challenged climate scientists for decades. Although the Noachian shows the greatest evidence of surface alteration by water, the sun was about 25\% less luminous at the time. Climate models have generally been unable to explain warm conditions on Mars under a faint young sun if CO$_2$ and H$_2$O are the only available greenhouse gases \citep{kasting1991co2}. This is because of the ``maximum greenhouse effect'' for an atmosphere dominated by CO$_2$, where the effects of greenhouse warming are balanced by the loss of incident radiation from Rayleigh scattering \citep{kasting1991co2,kopparapu2013habitable}; simply adding more CO$_2$ into an atmosphere is insufficient to warm early Mars. Even three-dimensional general circulation climate models, with seasonal and obliquity effects, are unable to easily yield warm and wet conditions for early Mars \citep[e.g.,][]{forget20133d,wordsworth2013global,wordsworth2015comparison}. 

One approach to this problem has been to invoke additional greenhouse gases. Hydrogen, H$_2$, has been suggested as an important greenhouse gas for extending the traditional concept of the liquid water habitable zone \citep{pierrehumbert2011hydrogen,wordsworth2013hydrogen} and explaining habitable conditions on early Mars \citep{ramirez2014warming,ramirez2017warmer,wordsworth2017transient}. A study by \citep{ramirez2014warming} demonstrated that Mars could have sustained above-freezing temperatures with a H$_2$ mixing ratio of 5\% or more, in addition to the greenhouse warming by several bars of CO$_2$. \rev{The composition of the early martian atmosphere remains uncertain, but factors such as the surface pressure \citep{gaillard2014theoretical} and the mantle redox state \citep{brachmann2025distinct} can constrain the type of atmosphere likely to form from mantle outgassing. High pressure atmospheres tend to be CO$_2$-dominated, while lower-pressure atmospheres and reduced mantle conditions can favor the outgassing of nitrogen species and other lighter species that are prone to atmospheric escape. If early Mars had a hydrogen greenhouse, then the} primary source of H$_2$ would be outgassing from a reduced mantle or serpentinization of ultramafic crust \citep{ramirez2014warming,batalha2015testing}, although impacts \citep{haberle2019impact} and crustal hydration \citep{adams2025episodic} have also been suggested as sources of H$_2$. Subsequent studies have even improved the viability of H$_2$O-CO$_2$-H$_2$ greenhouse warming on early Mars by considering the effects of collision-induced absorption in CO$_2$-dominated atmospheres \citep{wordsworth2017transient,turbet2019far,godin2020collision}. The inclusion of these additional absorption bands allows climate models to produce warmer conditions for early Mars than previously obtained; for example, above-freezing conditions can be achieved in some models with only about 1\% H$_2$ in a 3 bar CO$_2$ atmosphere, or about 20\% H$_2$ in a 0.5 bar CO$_2$ atmosphere \citep{wordsworth2017transient,ramirez2017warmer}. Theoretical calculations by \citet{wordsworth2017transient} as well as laboratory measurements by \citet{turbet2019far} and \citet{godin2020collision} have also evaluated collision-induced absorption features with CH$_4$ and CO$_2$, which suggests that a more complex mixture of greenhouse gases could have brought early Mars to above-freezing conditions. Previous efforts have attempted to resolve the faint young sun using a H$_2$O-CO$_2$-CH$_4$ atmosphere, but the formation of a stratospheric organic haze layer limits the amount of warming from CH$_4$ alone \citep{haqq2008revised,ramirez2014warming,arney2016pale,sauterey2022early}. Still, CH$_4$ remains an option to consider, perhaps in a H$_2$O-CO$_2$-H$_2$-CH$_4$ mixture. Although SO$_2$ has been suggested as a possible greenhouse gas on early Mars, SO$_2$ is expected to rain out or photolyze into stratospheric aeorosol and thus yield negligible warming \citep{tian2010photochemical,halevy2014episodic,batalha2015testing,kerber2015sulfur}.

Transient warming is another possible explanation for fluvial features on Mars, which suggests that Mars was primarily glacial but underwent one or several brief periods of warmth. One transient warming hypothesis is that impacts during the Late Heavy Bombardment created dense steam atmospheres, which then rained out and carved the valley networks \citep[e.g.,][]{segura2012impact}. Although this idea has been critiqued as insufficient to generate fluvial features in such a short time \citep[e.g.,][]{hoke2011formation}, the period of warming could potentially be prolonged through additional warming by cirrus clouds \citep{urata2013new,ramirez2017could} or other mechanisms. Another scenario for the climate of early Mars is the late Noachian icy highlands hypothesis \citep{Head_Marchant_2014,PALUMBO2018261,bishop2018surface,kite2019geologic}, which suggests that Mars experienced infrequent and regional exposure to warm temperatures. Drawing upon the Antarctic McMurdo Dry Valleys, an otherwise cold Mars could sustain seasonal meltwater runoff underneath a surface of thin ice. Impacts on an icy Mars might provide deviations from these cold conditions but would be unlikely to generate sustained warm periods that lead to the surface flow of water. The icy highlands hypothesis remains a viable alternative for the climate of Mars, which can maintain consistency with many geologic constraints without speculating on the availability of exotic greenhouse gases such as H$_2$ or CH$_4$.

Yet another transient warming hypothesis is that early Mars experienced climate ``limit cycles'' between prolonged periods of global glaciation and punctuated episodes of melting. Terrestrial planets with a low CO$_2$ outgassing rate or low incident stellar insolation are susceptible to fall into repeated cycles of global glaciation and deglaciation as a result of the dependence of CO$_2$ weathering on temperature and partial pressure of CO$_2$ \citep{tajika2003faint,kadoya2014conditions,kadoya2015evolutionary,kadoya2016evolutionary,menou2015climate}. When such a planet is in a glacial state, it is able to build up a sufficiently dense CO$_2$ atmosphere that eventually induces melting; however, the consumption of CO$_2$ by weathering on the deglaciated planet causes a reduction in greenhouse effect, which then plummets the planet back into glaciation. The onset of limit cycles for planets within the liquid water habitable zone depends on the volcanic outgassing rate as well as the stellar effective flux \citep{haqq2016limit}. Climate modeling studies have examined the possibility that limit cycles could explain the presence of fluvial features on Mars: a H$_2$O-CO$_2$ greenhouse is insufficient to warm early Mars and thus is insufficient to drive limit cycles, but the addition of H$_2$ allows both stable and cycling solutions to be obtained \citep{batalha2016climate,batalha2018reply,hayworth2020warming}. Limit cycling on early Mars depends upon the rates of CO$_2$ and H$_2$ outgassing, which must outpace loss from weathering (for CO$_2$) or escape to space (for H$_2$). Energy balance climate calculations by \citet{batalha2016climate} and \citet{hayworth2020warming} showed that limit cycles can be moderate, absent, or rapid, depending on the CO$_2$ and H$_2$ outgassing rates. Limit cycles remain an attractive solution for early Mars that combines greenhouse warming by several constituents with the possibility that warm phases of Mars were transient and punctuated by longer periods of glaciation. Even if Mars remained cold for much of its history, limit cycles could have warmed the planet for long enough to carve the valley networks.

Greenhouse gases, impacts, and limit cycles all remain possible explanations for a warm and wet period on early Mars, whether continuous or transient. \citet{kite2019geologic} described this trove of possible ideas as ``an embarrassment of riches'' that deserve further attention. Although one of these mechanisms might be able to provide a plausible explanation for the climate of early Mars, a combination of these factors is more likely able to explain the presence of at least two distinct river-forming periods over the history of Mars \citep{kite2019geologic}. A feature lacking from most climate studies of early Mars is explicit simulation of the long-term climate history of the planet. Most studies focus on establishing a steady-state warm scenario that applies to early Mars conditions, but few (if any) calculate climate trajectories beginning in the Noachian, transitioning into the Hesperian and eventually the dry Amazonian. Such long model integrations can be difficult with general circulation models, but simpler one-dimensional models can be adapted for this purpose. 

This study employs an energy balance climate model to calculate a \revv{schematic} long-term climate trajectory of Mars, which demonstrates that a combination of greenhouse warming and episodic limit cycling can effectively explain the timing of fluvial features. Such a hypothetical climate history remains consistent with available data, although it does not purport to be the only such trajectory that could exist for Mars. Instead, the purpose of this study is to show the feasibility of episodic warming based on limit cycling, which can provide a basis for future efforts at modeling or interpreting the combination of climate forcings that occurred across the history of Mars.

\section{Model Description}

The calculations in this study use a latitudinal energy balance model (EBM) to calculate steady-state climates and time evolution climate trajectories for terrestrial planets like Earth and Mars. This class of EBMs has a long history of application to problems of understanding ice-albedo feedback and hysteresis in terrestrial planets \citep[see, e.g.,][]{north2017energy}, with applications ranging from understanding snowball Earth episodes on Earth to exploring possible warm and wet conditions on early Mars. The specific model used here is the habitable energy balance model for exoplanet observations \citep[HEXTOR;][]{haqq2022energy}, which calculates the latitudinal profile of temperature $T$ using the diffusive EBM equation:
\begin{equation}
C\frac{\partial T}{\partial t}=S\left(1-\alpha\right)-F+\frac{\partial}{\partial x}\left[D\left(1-x^{2}\right)\frac{\partial T}{\partial x}\right],\label{eq:EBMfull}
\end{equation}
where the solar insolation $S$ is a function of time and zenith angle (to represent the seasonal cycle); the dimension $x$ is the sine of latitude; the diffusive parameter $D$ represents meridional energy transport; the thermal heat capacity $C$ depends on temperature and the land-ocean distribution; the outgoing infrared radiative flux $F$ depends on temperature \rev{and} the partial pressure of CO$_2$; and the planetary albedo $\alpha$ depends on temperature, zenith angle, the partial pressure of CO$_2$, and the surface albedo. \revminor{For the limit of a point Earth with no latitudinal divisions, the diffusive term vanishes and Eq. (\ref{eq:EBMfull}) reduces to show that the change in temperature depends on a balance between incoming (downward-directed and positive-signed) solar radiation and outgoing (upward directed and negative-signed) infrared radiation.}

\begin{figure}[ht!]
\centerline{\includegraphics[width=5.5in]{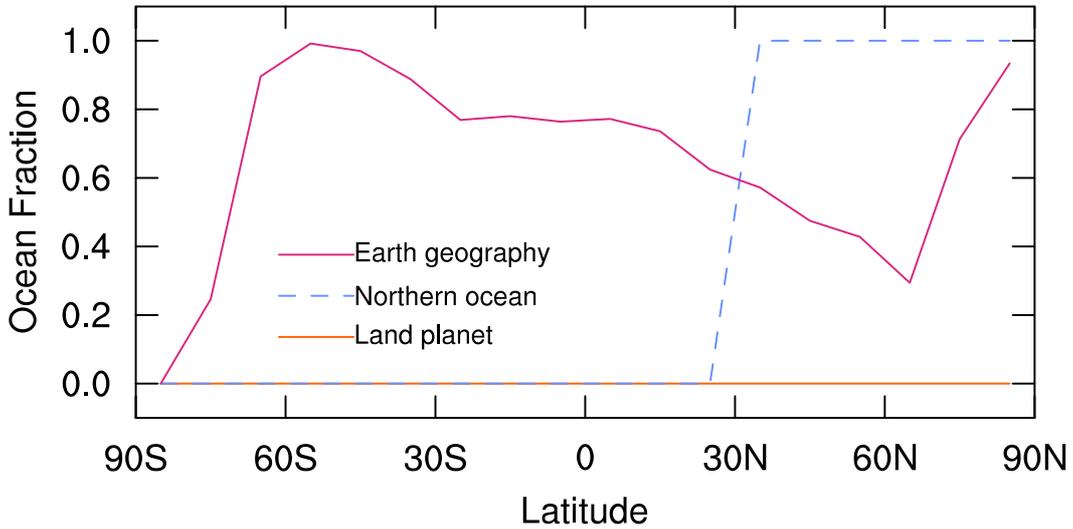}}
\caption{\rev{The present Earth (magenta), northern ocean (dashed blue), and land planet (orange) land-ocean distributions considered in this study are shown as ocean fraction versus latitude.}\label{fig:geography}}
\end{figure}

The model configuration is identical to the detailed description provided by \citet{haqq2022energy}; this includes a fixed value of $D=0.38$\,W\,m$^{-2}$\,K$^{-1}$; surface albedo is set to 0.3 for unfrozen land, 0.663 for frozen land, and following a Frensel reflection equation for ocean; heat capacity set to $5.25\times10^6$\,J\,m$^{-2}$\,K$^{-1}$ over unfrozen land, twice this value over ice, and 40 times this value over ocean; and a fractional land-ocean distribution defined at each of the 18 equally-spaced latitudinal bands\rev{. This study will consider three possible land-ocean distributions that correspond} to Earth-like geography, a land planet, \rev{and a northern ocean} (Fig. \ref{fig:geography}). The model uses a finite difference scheme with a time step of $\Delta t =12$\,hr to calculate the change in temperature over a complete orbit. The model assumes an atmosphere of 1\,bar N$_2$, with the contributions from CO$_2$ added to this background. The values of $\alpha$ and $F$ are both calculated using a lookup table, which was tabulated using thousands of calculations using the radiative-convective climate model of \citet{kopparapu2013habitable}. Further details regarding the implementation of the radiative transfer lookup table or other aspects of the model configuration are described by \citet{haqq2022energy}.

This study also considers the role of additional greenhouse gas forcing beyond CO$_2$ alone. The HEXTOR lookup tables for $F$ only account for greenhouse warming by CO$_2$, so the contributions from other sources of warming are accounted for with a generic term $F_{add}$ that is added to Eq. (\ref{eq:EBMfull}) as shown:
\begin{equation}
C\frac{\partial T}{\partial t}=S\left(1-\alpha\right)-F+F_{add}+\frac{\partial}{\partial x}\left[D\left(1-x^{2}\right)\frac{\partial T}{\partial x}\right].\label{eq:EBMfull_Fadd}
\end{equation}
Such a simplified representation is useful in this study because of the range of possible additional constituents that could have contributed to greenhouse warming on Mars, such as H$_2$, CH$_4$, and other species. Rather than explore the dependence of possible early Mars climates on the abundances or mixtures of specific gases, this study takes a more generic approach by examining the dependence of climate states on the additional flux of warming provided by an arbitrary set of additional gases. The term $F_{add}$ could even represent other sources of warming (such as warming by impacts), although the use in this study will generally consider $F_{add}$ as a fixed contribution that remains steady in time.

The long-term carbon cycle, known as the carbonate-silicate cycle, follows the same functional form as used by \citet{haqq2016limit}:
\begin{equation}
\frac{d}{d\tau}(p\text{CO}_{2})=V-W\label{eq:co2evo},
\end{equation}
and 
\begin{equation}
\frac{W}{W_{\mathTerra}}=\left(\frac{p\text{CO}_{2}}{p_{\mathTerra}}\right)^{\beta}e^{k_{act}(T-288)}\left[1+k_{run}(T-288)\right]^{0.65}\label{eq:weathering},
\end{equation}
where $p\text{CO}_{2}$ is the carbon dioxide partial pressure, $V$ is the volcanic outgassing rate, $W$ is the silicate weathering rate, $p_{\mathTerra}$ represents the long-term balance between atmospheric and soil CO$_{2}$, $k_{act}=0.09$ is an activation energy, $k_{run}=0.045$ is a runoff efficiency factor, and $\beta = 0.5$ is a factor that determines the rate of $p\text{CO}_{2}$ evolution. The time variable $\tau$ represents a slower timestep at which $p\text{CO}_{2}$ evolves, which is contrasted with the timestep $t$ in Eq. (\ref{eq:EBMfull}) for calculating temperature over a full orbit. Further details on this implementation of the carbonate-silicate cycle are described by \citet{haqq2016limit}.

\revv{The two most influential factors in determining whether or not a planet resides in a limit cycling regime are the volcanic outgassing rate, $V$, and the partial pressure of CO$_2$ in soil, $p_{\mathTerra}$ (see, e.g., \citet{haqq2016limit}). This study does not attempt a full parameter space exploration of the dependence of limit cycling on these two parameters, as this parameter space has already been mapped out by \citet{haqq2016limit} (see their Fig. 2) using a prior version of the HEXTOR model that gives similar results. Instead, this study draws upon these prior calculations to select parameters that optimize the presence of limit cycles on a terrestrial planet.} The calculations in this study assume \revvv{$V=7$\,bar\,Gyr$^{-1}$,} which corresponds to \revvv{14\% (one-seventh)} the outgassing rate on a tectonically active planet like modern Earth \citep{haqq2016limit}. \revvv{The volcanic outgassing rate throughout the history of Mars remains difficult to constrain, but it is perhaps reasonable that the volcanic outgassing rate of small terrestrial planets should (to first-order) scale with planetary mass \citep[e.g.,][]{dorn2018outgassing}.} \revv{The value of $p_{\mathTerra}$ may be the most difficult to constrain, and various authors \citep[e.g.,][]{menou2015climate,kadoya2014conditions,haqq2016limit} have made different assumptions for this parameter. The study by \citet{haqq2016limit} noted that $p_{\mathTerra}$ represents a long-term balance between atmospheric and soil CO$_2$ (with other factors held constant), which on Earth today includes contributions from root respiration by vascular plants that increase soil CO$_2$. Studies by \citet{menou2015climate} and \citet{kadoya2014conditions} used a value of $p_{\mathTerra} \sim 10^{-4}$\,bar, which corresponds to present-day atmospheric CO$_2$ levels; however, \citet{haqq2016limit} suggested that this factor should be enhanced to $p_{\mathTerra} \sim 10^{-2}$\,bar to account for the soil sequestration of CO$_2$ by land plants on Earth.} \revvv{The partial pressure of CO$_2$ on Mars today is about $6\times10^{-3}$\,bar, and may have been higher in the past, so this represents a minimum value of $p_{\mathTerra}$ for Mars. Given that this is only a minimum, this study will use the value of $p_{\mathTerra} \sim 10^{-2}$\,bar suggested by \citet{haqq2016limit}.}

\rev{The choice of parameters for the outgassing and weathering expressions are designed to place the planet in a regime that is prone to limit cycling. This is based on the results of prior calculations by \citet{haqq2016limit} that showed limit cycling for a planet with \revvv{$p_{\mathTerra}=10^{-2}$\,bar and $V=7$\,bar\,Gyr$^{-1}$,} assuming a solar flux of $S/S_0 = 0.7$. But the same set of calculations show that a planet would reside in a stable, non-cycling region at $V>50$\,bar\,Gyr$^{-1}$. The selection of fixed values for $p_{\mathTerra}$ and $V$ in this study is intended to focus the investigation on situations in which a planet is prone to rapid limit cycling. Subsequent work could further investigate plausible climate trajectories for Mars by varying these parameters to explore regions where Mars has less rapid limit cycling or resides in a stable climate. This study intentionally focuses on a case where Mars is prone to limit cycling for its entire modeled history, but improved climate trajectories for Mars could include time-dependent variation in $p_{\mathTerra}$ and $V$---ideally, using values with improved constraints from observations or geologic records on Mars.}

\rev{This model of the carbonate-silicate cycle is also based on parameterizations of Earth, a tectonically active planet, so the application of this model to Mars implies that similar tectonic mechanisms were operating on Mars during much of its history. Some interpretations of available martian data suggest that periods of the Noachian and Hesperian may have featured active plate tectonics \citep[e.g.,][]{sleep1994martian}, which would be consistent with the assumption in this study of continuous volcanic activity during this duration. However, Mars could have instead been in a stagnant lid tectonic regime during its entire history \citep[e.g.,][]{breuer2003early}, which would limit the availability of outgassed CO$_2$ over long periods of time. One possibility is that point volcanism could continue to release CO$_2$ from the mantle to the atmosphere on a stagnant lid planet; \citet{hayworth2020warming} estimated that point volcanism could remain constant for longer than the age of Mars without depleting the carbon reservoir. Weathering may also have been different on a stagnant lid planet, due to the lack of reweatherable material being supplied to the surface, which can cause dense CO$_2$ atmospheres to form. Some models have suggested the possibility that stagnant lid planets could sustain carbon cycling for a finite duration of 1--5\,Gyr \citep{foley2018carbon}, which could be an alternative approach for driving limit cycles during the Noachian and Hesperian periods. The Earth-centric weathering parameterization will be used in the calculations that follow, keeping in mind that a stagnant lid Mars could be comparable in some instances.}

\revv{Finally, it is important to note that the weathering model (Eq. (\ref{eq:weathering})) was developed by \citet{berner2001geocarb} for application to an Earth-like planet with large volumes of liquid water oceans available throughout the planet's history. Applying this model to Mars therefore assumes that any weathering processes that occurred on Mars were comparably ``Earth-like,'' which may not necessarily have been the case if Mars oscillated between wet and dry conditions. The parameter $k_{run}$ is a ``coefficient expressing the effect of temperature on global river runoff'' \citep{berner2001geocarb}, which was determined from the use of general circulation climate models of paleo-Earth. The study by \citet{berner2001geocarb} assigned the value of $k_{run} = 0.045$ for colder periods in Earth's history (from 340--260\,Ma and 40--0\,Ma) and value of $k_{run} = 0.025$. One extension of this model could be to utilize general circulation models for past and present Mars to better constrain values of $k_{run}$ during wet and dry conditions; however, \citet{berner2001geocarb} also noted that changes in runoff show a relatively low sensitivity to CO$_2$. By contrast, CO$_2$ is much more sensitive to changes in the activation energy defined as the coefficient $k_{act} = {E}/{R T^2}$, where $E$ is the dissolution activation energy and $R$ is the gas constant \citep{berner2001geocarb}. Values of $k_{act}$ are based on field studies; the value of $k_{act} = 0.09$ was preferred by \citet{berner2001geocarb}, although it was also noted that other values ranging from 0.06 to 0.135 have been reported. For Mars, factors such as temperature oscillations between warm and cold climates as well as differences in mineralogical and atmospheric composition may all contribute to changes in the values of $k_{act}$, which would drive higher or lower values of CO$_2$ than would be obtained with fixed value based on Earth observations. Such exploration of the weathering model is beyond the scope of the present study, but it is worth emphasizing the extent to which the weathering model depends on Earth-based assumptions.}

\section{Limit Cycles on Earth and Mars}

Terrestrial planets with active carbonate-silicate cycles can regulate their long-term temperature through a balance of CO$_2$ accumulation in the atmosphere from volcanic outgassing and draw-down of CO$_2$ by weathering (Eq. (\ref{eq:co2evo})). The outermost edge of the liquid water habitable zone \citep{kasting1993habitable,kopparapu2013habitable} is defined as the farthest orbital distance at which an Earth-like planet with an active carbonate-silicate cycle can maintain a stable atmosphere based on CO$_2$ greenhouse warming. A planet at a more distant orbit than Earth today would receive less insolation, thereby causing the climate to be cooler; this in turn reduces the rate of silicate weathering and allows more CO$_2$ to accumulate in the atmosphere. This stabilizing feedback provides a way for such planets to retain warm atmospheres at orbits as far as Mars today, as long as the planet remains tectonically active. Beyond the outer edge of the habitable zone, even a tectonically active planet would not be able to stay sufficiently warm from additional accumulation of CO$_2$, as the increase in Rayleigh scattering from such a dense CO$_2$ atmosphere would exert net cooling and cause the planet to freeze over into glacial conditions. 

However, the weathering rate on terrestrial planets depends not only on temperature but also on the CO$_2$ partial pressure, as noted by \citet{menou2015climate} and shown in Eq. (\ref{eq:weathering}). The accumulation of CO$_2$ itself would also cause increases in weathering rate along with increases in temperature, and likewise a decrease in CO$_2$ abundance would slow the rate of weathering along with decreases in temperature. This can lead to situations in which planets that receive less insolation than Earth today are prone to ``limit cycling'' climates that oscillate between warm and glacial states: such a planet in a glacial state will accumulate atmospheric CO$_2$ up to the threshold for deglaciation, but the increased rate of weathering after deglaciation causes a rapid draw down of CO$_2$ that returns the planet to its glacial state. Although a planet like modern Earth should not experience such limit cycles, limit cycles could have occurred during periods of Earth's early history when the sun was fainter. In general, planets orbiting toward the outer edge of the liquid water habitable zone around G-dwarf stars would be prone to limit cycling, particularly if CO$_2$ outgassing rates are less than Earth today \citep{haqq2016limit}.

Limit cycles could also potentially explain the presence of fluvial features on Mars. The longstanding problem of explaining warm and wet early Mars requires invoking some additional warming mechanism beyond CO$_2$ alone, and the addition of other greenhouse gases has been suggested as a possibility. The combination of greenhouse warming by CO$_2$ and H$_2$ could allow for warm conditions to exist on Mars \citep{ramirez2014warming}, whether continuously or episodically. Limit cycling in such CO$_2$-H$_2$ atmospheres is one explanation for fluvial features on Mars, as variations in both the CO$_2$ and H$_2$ outgassing rates could have provided conditions at one or more points in the history of Mars to enable transient warm periods that persisted for up to $\sim$10\,Myr before returning to glaciation \citep{batalha2016climate,batalha2018reply,hayworth2020warming}. Limit cycling may have only occurred sporadically during the evolution of Mars, but such events could conceivably have been sufficiently frequent and of long enough duration to explain how fluvial features were generated on an otherwise cold and dry planet.

The susceptibility of a tectonically active terrestrial planet to limit cycling depends on factors that include the volcanic outgassing rate, the planetary obliquity, and any other greenhouse gas forcing beyond CO$_2$ and water vapor. In this study, the volcanic outgassing rate is fixed at \revminor{14\%} present-day Earth levels in order to enable a systematic exploration of the other two factors that contribute to limit cycling. The addition of other greenhouse gases beyond CO$_2$ is also abstracted as a generic parameter, $F_{add}$, that represents the infrared flux generated by any arbitrary combination of greenhouse cases (including H$_2$, CH$_4$, and other possibilities) that would enable early Mars to reach above-freezing conditions. \rev{Many atmospheric gases that could have provided additional greenhouse warming, such as H$_2$ and CH$_4$, would likely have varied over time due to hydrogen escape; in this sense, $F_{add}$ would ideally be represented as a time-varying quantity. However,} this study is not concerned with identifying any single gas or mixture of gases that would contribute to $F_{add}$ but instead is focused on understanding the limit cycling behavior of a terrestrial planet as $F_{add}$ and planetary obliquity vary. \rev{The choice of a time-invariant value for $F_{add}$ is intended to focus this study on the relationship of limit cycling on obliquity, but time-variance in $F_{add}$ could serve as another mechanism that could trigger limit cycling.}

\begin{figure}[ht!]
\centerline{\includegraphics[width=4.5in]{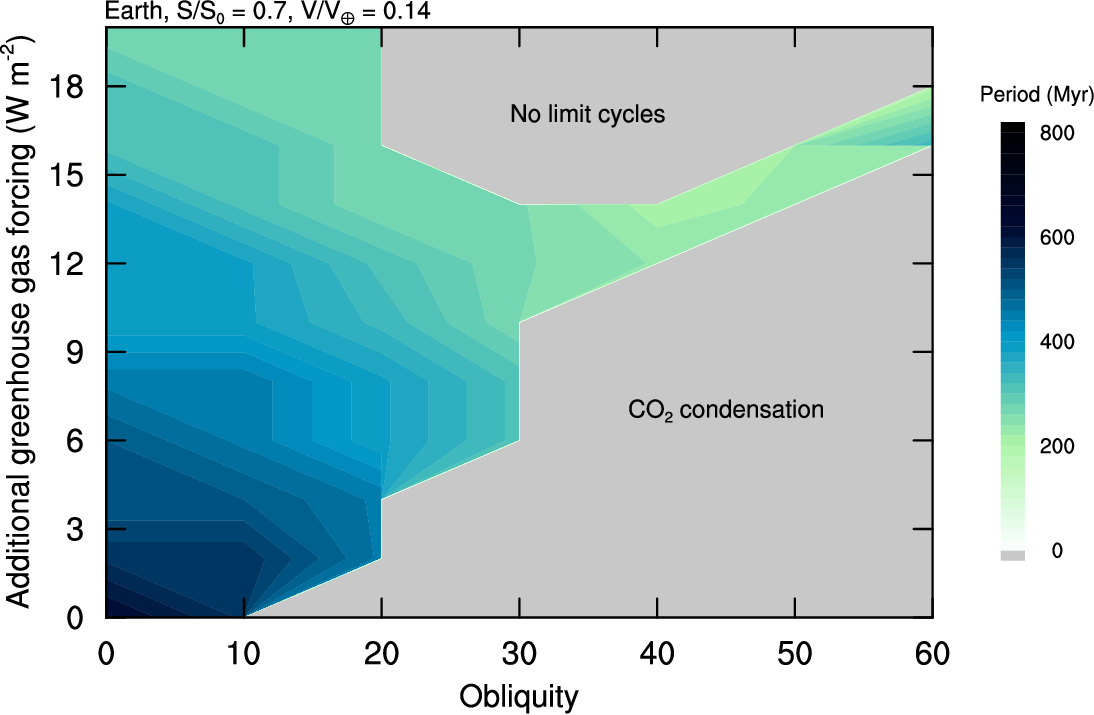}}
\caption{The \revvv{period} of limit cycle \revvv{events} (color contours) depends on planetary obliquity and the magnitude of any additional greenhouse gas warming $F_{\text{add}}$. Limit cycling is assumed to halt when CO$_2$ condensation occurs. \revvv{At high obliquity and large additional greenhouse forcing, climate remains warm and limit cycles do not occur.} Calculations are performed at an early Earth solar constant of $S/S_0=0.7$ for a 1\,bar N$_2$ terrestrial planet atmosphere with an active carbonate-silicate cycle, \revvv{14\%} present-day volcanic outgassing rates, and a land-ocean distribution that corresponds to present Earth.\label{fig:obliquity}}
\end{figure}

For \revvv{an} early \revvv{Earth-like planet}, the \revminor{period} of limit cycles as a function of obliquity and additional greenhouse gas forcing is shown in Figure \ref{fig:obliquity}. These calculations are performed at a solar constant of $S/S_0 = 0.7$ and a land-ocean distribution that corresponds to present-day Earth, with a 1\,bar N$_2$ atmosphere and the CO$_2$ partial pressure determined by the active carbonate-silicate cycle. These calculations indicate that limit cycling is possible for early Earth at low values of obliquity when $F_{add} = 0$, with CO$_2$ condensation occurring beyond an obliquity of 10$^{\circ}$. As $F_{add}$ increases, the range of obliquity values at which limit cycling occurs also expands, which permits warm climates to occupy a wider range of the parameter space before CO$_2$ condensation occurs. This behavior is expected, as the contribution of additional infrared warming to the surface helps to warm the poles and allow the planet to maintain warm conditions at high obliquity while staying above the CO$_2$ condensation threshold. Likewise, the \revvv{period} of limit cycling \revvv{events (in Myr)} \revminor{decreases} with greater values of $F_{add}$; this indicates the contribution of additional greenhouse gas warming to lowering the threshold for deglaciation, thereby resulting in more frequent and shorter duration limit cycling as CO$_2$ is outgassed and drawn down. \revvv{Limit cycles no longer occur at high obliquity and high $F_{add}$, with the planet remaining in a stable steady-state climate.} The calculations shown in Figure \ref{fig:obliquity} are illustrative of the conditions in an early Earth setting that would have resulted in limit cycling, although they are not necessarily intended to demonstrate that limit cycling actually occurred at this \revminor{time} in Earth's history. Instead, these calculations provide a benchmark for understanding how similar changes in obliquity and additional greenhouse gas forcing would contribute to limit cycling behavior on early Mars.

\begin{figure}[ht!]
\centerline{\includegraphics[width=5.5in]{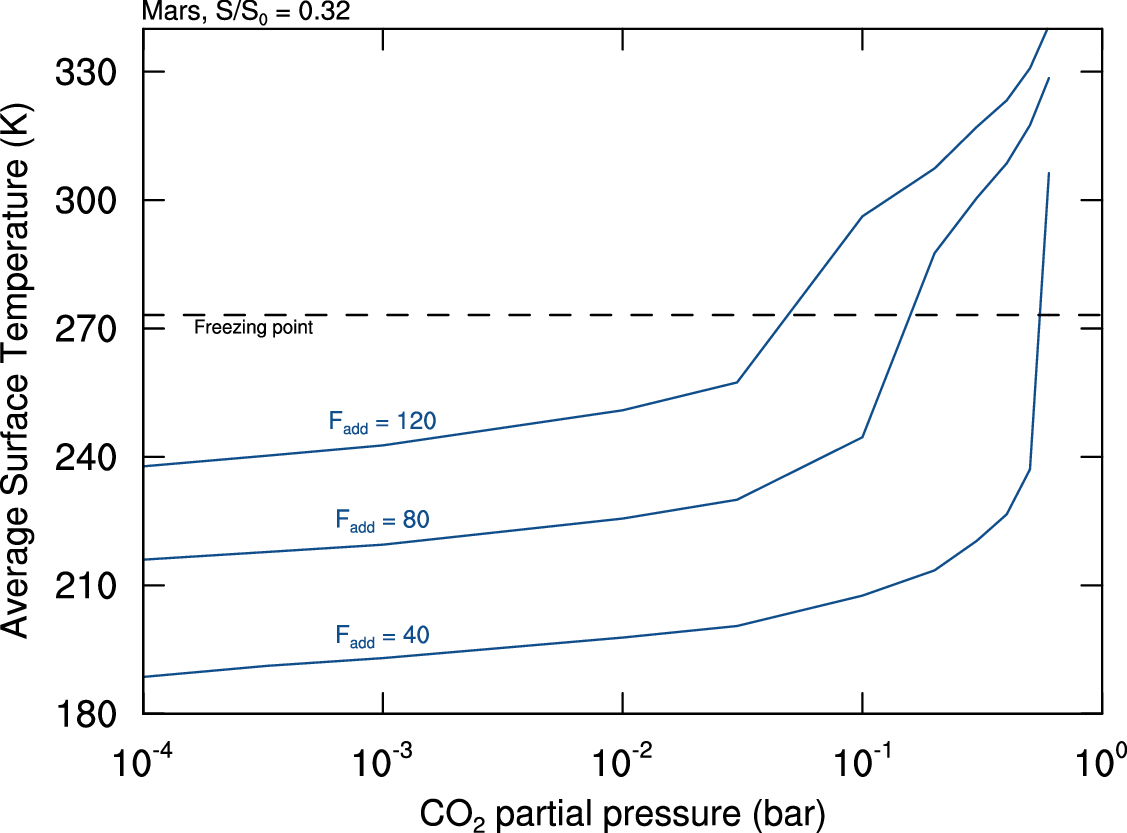}}
\caption{At an early Mars solar constant of $S/S_0=0.32$, additional greenhouse gas warming is required for a CO$_2$-rich atmosphere to reach an average temperature above the freezing point. Calculations show average surface temperature as a function of CO$_{2}$ partial pressure for a 1\,bar N$_2$ terrestrial planet and additional greenhouse gas warming $F_{\text{add}}$, at 40, 80, and 120\,W\,m$^{-2}$. Cases assume a land-ocean distribution that corresponds to present Earth.\label{fig:tempco2}}
\end{figure}

Limit cycling, like other explanations for martian fluvial features, could not occur on early Mars without an additional source of surface warming. No amount of CO$_2$ would be sufficient to allow early Mars to reach above-freezing temperatures, given the decreased insolation of the faint young sun \citep[e.g.,][]{kasting1991co2,ramirez2014warming}. This behavior is illustrated in Figure \ref{fig:tempco2}, which shows the average surface temperature calculated for early Mars as a function of CO$_2$ partial pressure and at three different values of $F_{add}$. These calculations are performed at a solar constant of $S/S_0 = 0.32$, appropriate for early Mars, with a land-ocean distribution that corresponds to present-day Earth and a a 1\,bar N$_2$ atmosphere in addition to the CO$_2$ partial pressure. These results demonstrate that even a CO$_2$-dominated atmosphere requires at least 40\,W\,m$^{-2}$ of additional warming to exceed the freezing point of water. The additional warming could take the form of any greenhouse gas, or even could be other forms of warming such as impacts, but some sort of additional warming mechanism is needed for either a steady-state warm early Mars or limit cycling early Mars.

\begin{figure}[ht!]
\centerline{\includegraphics[width=5.5in]{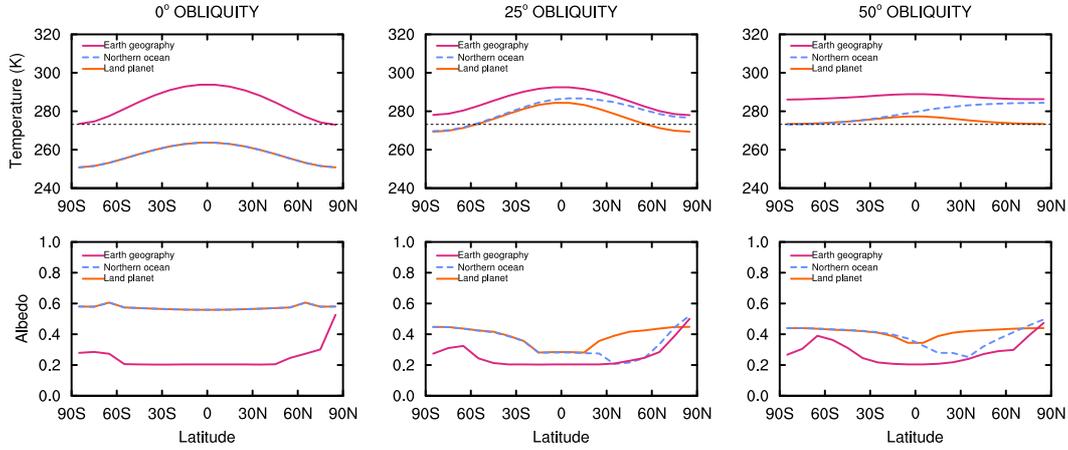}}
\caption{Latitudinal profiles of average temperature (top row) and planetary albedo (bottom row) for a 1\,bar N$_2$ terrestrial planet with a 0.2\,bar CO$_{2}$ partial pressure, an early Mars solar constant of $S/S_0=0.32$, and an additional $F_{\text{add}}=80$\,W\,m$^{-2}$ of greenhouse warming. Cases are shown at obliquities of 0$^{\circ}$ (left column), 25$^{\circ}$ (middle column), and 50$^{\circ}$ (right column), with a land-ocean distribution that corresponds to present Earth (magenta), a northern ocean (\rev{dashed} blue), and a land planet (orange). \rev{The dashed black line indicates the freezing point of water.}\label{fig:marstempalb}}
\end{figure}

The influence of obliquity and the geographical land-ocean distribution on the steady-state temperature and albedo of early Mars climates is shown in Figure \ref{fig:marstempalb}. These calculations are performed at a solar constant of $S/S_0 = 0.32$ with a 1\,bar N$_2$ atmosphere and a 0.2\,bar CO$_2$ partial pressure, as well as $F_{add}=80$\,W\,m$^{-2}$. The three obliquities of 0$^{\circ}$, 25$^{\circ}$, and 50$^{\circ}$ are selected to span the approximate range of obliquity values that Mars has experienced during its long-term evolution \citep{laskar1993chaotic,laskar2004long}. The land-ocean distribution corresponding to Earth geography is the same as was used in the calculations shown in Figures. \ref{fig:obliquity} and \ref{fig:tempco2}, and the land planet distribution corresponds to complete continental coverage with no oceans but that still allows for the formation of snow and evaporation of groundwater. (Comparable behavior for these two land-ocean distributions is shown in prior calculations with this model for present-day Earth by \citet{haqq2022energy}.) The northern ocean distribution is intended to represent a martian configuration with the northernmost 25\% of the planet covered by an ocean and the remainder of the planet as land. The land planet distribution shows complete symmetry for the latitudinal profiles of temperature and albedo, with maximum temperature and minimum albedo at the equator for all cases. The Earth geography distribution shows a \rev{symmetric latitudinal temperature profile but an asymmetric albedo profile; the higher albedo} at the northern pole \rev{corresponds to an ocean-dominated region that remains frozen much of the year. Such temperature differences from albedo and others that might arise from Earth-like geography (such as temperature asymmetries due to the Antarctic continent) are muted as a result of the dense 0.2\,bar CO$_2$ atmosphere along with the additional $F_{\text{add}}=80$\,W\,m$^{-2}$ of greenhouse warming.} The northern ocean distribution shows lower albedo and higher temperature values in the northern hemisphere. For all cases, these latitudinal differences are more pronounced at higher values of obliquity.

\begin{figure}[ht!]
\centerline{\includegraphics[width=5.5in]{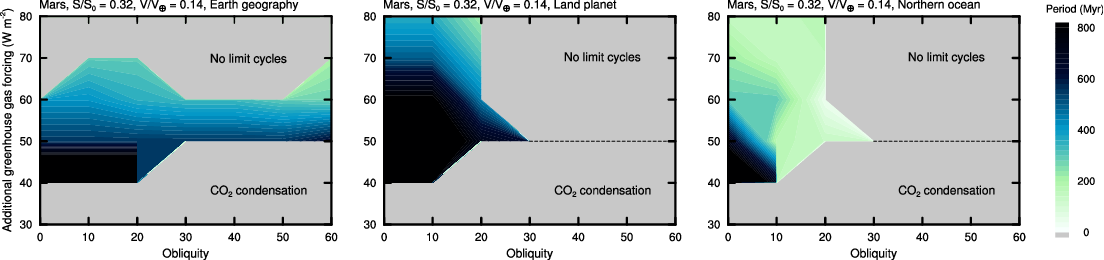}}
\caption{The relationship between the \revvv{period} of limit cycle \revvv{events} (color contours) and additional greenhouse warming depends on the land-ocean distribution. Calculations are performed at an early Mars solar constant of $S/S_0=0.32$ for a 1\,bar N$_2$ terrestrial planet atmosphere with an active carbonate-silicate cycle, \revvv{14\%} present-day volcanic outgassing rates, and land-ocean distributions that correspond to present Earth (left), a land planet (middle), and a northern ocean (right). All cases show a region at high obliquity and large additional greenhouse forcing where climate remains warm and limit cycles do not occur. At high obliquity and small additional greenhouse forcing, limit cycling is assumed to halt as CO$_2$ condensation occurs.\label{fig:obliquitylimcycle}}
\end{figure}

For early Mars, the \revminor{period} of limit cycles as a function of obliquity and additional greenhouse gas forcing is shown in Figure \ref{fig:obliquitylimcycle}. These calculations are performed at a solar constant of $S/S_0 = 0.32$ with a 1\,bar N$_2$ atmosphere and the CO$_2$ partial pressure determined by the active carbonate-silicate cycle. The three panels show how the limit cycling region in this parameter space is dependent on the assumed land-ocean distribution. All cases require at least $F_{add}=40$\,W\,m$^{-2}$ to avoid CO$_2$ condensation. \rev{(For comparison, a large volcanic eruption could cause a change of 5--10\,W\,m$^{-2}$, while the addition of 20\% H$_2$ to a 2-bar CO$_2$ atmosphere would correspond to about 20\,W\,m$^{-2}$ of additional warming \citep{ramirez2014warming}; a value of $F_{add}=40$\,W\,m$^{-2}$ could correspond to a scenario with significant warming by H$_2$ as well as other gases such as CH$_4$, perhaps with contributions from volcanism.)} The Earth geography distribution allows for limit cycles to occur across the full range of obliquity values from 0$^{\circ}$ to 60$^{\circ}$, although the limit cycling region transitions to a stable climate regime at larger values of $F_{add}$. Both the land planet distribution and the northern ocean distribution only show limit cycles up to obliquity values of 20$^{\circ}$--30$^{\circ}$, with a stable climate regime at higher obliquity values. This behavior indicates that high obliquity climates (with large values of $F_{add}$) remain ice-free due to the extreme seasonal warming at the poles. However, limit cycling region for the northern ocean distribution shows much more rapid freeze-thaw cycles than the land planet, due to its greater susceptibility to ice-albedo feedback in the northern hemisphere. These calculations illustrate the parameter space under which limit cycling could have occurred on early Mars, although these calculations alone do not necessarily imply that limit cycling occurred at any particular point in the history of Mars.

The model calculations that have been discussed in this section are intended to demonstrate that limit cycling is a feature of terrestrial planets that could conceivably apply to Mars, and that any limit cycling behavior depends on the planet's obliquity while also requiring an additional source of warming to counteract the faint young sun. The land-ocean distribution likewise can change the region of parameter space susceptible to limit cycling; other configurations for this surface distribution could be explored in subsequent calculations, but the comparison between present-day Earth, a land planet, and a northern ocean will suffice for the purposes of this study. The goal of this study is not to conclusively demonstrate that limit cycles must have occurred on Mars but instead to suggest that limit cycles are a plausible hypothesis to explain the timing of fluvial events in the history of Mars. The calculations shown so far have focused on steady-state or time average model results, but the next section will apply this model configuration to the time evolution of the climate of Mars.

\section{A Climate History of Mars}

Episodic limit cycling during the Noachian and Hesperian periods provides one hypothetical explanation for the formation of fluvial features on Mars. The model calculations presented in this section draw upon the exploration of limit cycling behavior in the previous section to generate a time-forward trajectory for a plausible climate history of Mars. This exercise is intended to demonstrate the viability of the episodic limit cycling hypothesis for Mars, although the specific trajectory shown is only an example of such a time-forward trajectory rather than an assertion of the most likely trajectory.

Changes in the solar luminosity with time are expected due to the ongoing evolution of the sun during its main sequence lifetime. The sun was about 30\% less luminous during the earliest periods of history on Earth and Mars, which only exacerbates the problem of explaining a warm and wet climate. For these time-forward calculations, changes in the luminosity of the sun follow the expression developed by \citet{gough1981solar}:
\begin{equation}
    L(t) = L_{\astrosun}\left[1+\frac{2}{5}\left(1-\frac{t}{t_{\astrosun}}\right)\right]^{-1},\label{eq:gough}  
\end{equation}
where $t_{\astrosun}=4.57$\,Gyr. The value of $L_{\astrosun}$ is the present-day solar luminosity, which in this case is at the orbital distance of Mars.

The obliquity of Mars also evolves chaotically over the planet's history, due to gravitational perturbations from Jupiter and the lack stabilization from a large moon. For these time-forward calculations, obliquity, $\Phi$ (in degrees), varies according to a prescribed sinusoidal function, which is intended to mimic the chaotic obliquity ranging from 0$^{\circ}$ to about 60$^{\circ}$ \citep[][]{laskar1993chaotic,laskar2004long} that Mars would have experienced during its history:
\begin{equation}
    \Phi = 25\sin\left(\frac{\pi t}{\omega}\right)+25,\label{eq:obliquityfunction}
\end{equation}
where $\omega = 450$\,Myr. This parameterized function for obliquity is admittedly an over-simplification, but it is impossible to fully recover a quantitative time series for the obliquity evolution of Mars across its entire history. \rev{This choice was made by inspecting the various trajectories calculated by \citet{laskar2004long} for the evolution of the obliquity of Mars; these solutions all show a short-timescale ($\sim$0.1 Myr) chaotic variation of about $\pm$10 degrees, but many of the solutions show a much longer periodicity for Mars to oscillate between the lowest (0 degrees) and highest (50--60 degrees) ranges of its obliquity. The simplified function in Eq. (\ref{eq:obliquityfunction}) is intended to capture this longer-term variation between obliquity extremes.} \revv{However, this approach neglects the contributions of shorter-term chaotic variations in obliquity on climate, which \citet{laskar2004long} showed can be as significant as the long-term variability. The chaotic nature of the obliquity of Mars makes deterministic simulation impossible beyond relatively recent time horizons, so} other approaches \revv{would need to} be implemented for representing \revv{such rapid and chaotic} changes in the obliquity of Mars over time, such as a stochastic function \revv{and an ensemble of model results. The} simplified sinusoidal variation \revv{used in} this study \revv{is sufficient} to show that changes in obliquity can contribute to episodic limit cycling, \revv{but this also means that the resulting calculations cannot be considered a completely climate history of Mars, to the extent that short-term chaotic obliquity influences on climate are not included. Such exploration will be reserved for future work, while the present discussion will emphasize that these results are a ``schematic'' climate history of Mars at best rather than a ``complete'' climate history.}

\begin{figure}[ht!]
\centerline{\includegraphics[width=5.5in]{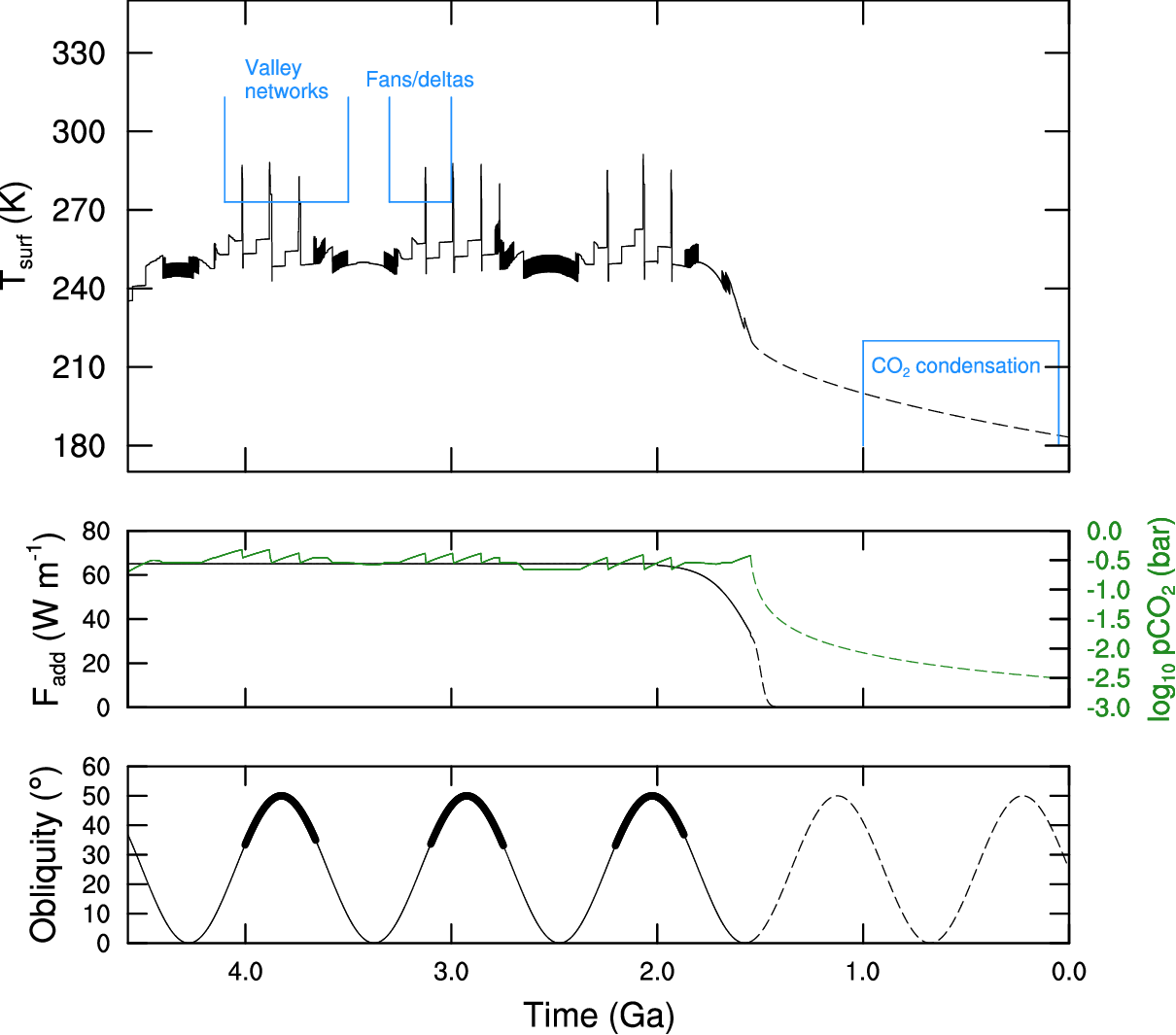}}
\caption{Time-forward model calculations demonstrate a \revv{schematic} trajectory for the climate history of Mars. Warm periods in the history of Mars during the formation of the valley networks and fans/deltas could correspond to episodic intervals of limit cycling \rev{(spikes in surface temperature)} alternating with periods of cooler climates \rev{(low variation in surface temperature)}. The solar constant evolves to represent the insolation from a faint young sun that brightens with time \citep{gough1981solar}. Planetary obliquity varies according to a prescribed sinusoidal function \revv{as a simplified representation of changes in} obliquity that Mars would have experienced. The model begins with glacial conditions, a 1\,bar N$_2$ terrestrial planet atmosphere with an active carbonate-silicate cycle, \revvv{14\%} present-day volcanic outgassing rates, and a northern ocean land-ocean distribution. A fixed value of \revvv{$F_{\text{add}}=65$\,W\,m$^{-2}$} represents additional greenhouse warming that could have been obtained from volcanic activity. The model integration continues from 4.3\,Ga to 1.8\,Ga, which corresponds to the duration of active volcanism on Mars; volcanism ceases after this point, which causes the planet to drop to below-freezing temperatures and eventually reach the CO$_2$ condensation threshold. Dashed lines indicate extrapolations from the end state of the simulated climate. \revv{Bold regions of the obliquity curve correspond to periods of limit cycling.}\label{fig:tempseries}}
\end{figure}

The \revv{schematic} time-forward trajectory of temperature, CO$_2$ partial pressure, \rev{and obliquity} calculated with the model across the entire history of Mars is shown in Figure \ref{fig:tempseries}. These results show explicit model calculations from 4.3\,Ga to 1.8\,Ga at a step interval of 100,000 years (smaller increments as low as 100 years were also attempted, but these did not change the results), which corresponds to the period of active volcanism on Mars during the Noachian and Hesperian. This interval includes a constant \revvv{$F_{add}=65$\,W\,m$^{-2}$} as additional greenhouse warming associated with volcanic outgassing of H$_2$ or other constituents. Solar forcing changes according to Eq. (\ref{eq:gough}), and obliquity varies according to Eq. (\ref{eq:obliquityfunction}). From 1.8\,Ga to the present, the cessation of volcanic activity causes the planet to plummet into a glacial state, in which silicate weathering halts, $F_{add}$ is removed, and CO$_2$ levels drop; explicit model calculations are not performed for this later time period, and Figure \ref{fig:tempseries} instead shows dashed curves along approximate trajectories as the planet eventually enters the CO$_2$ condensation regime. These calculations begin with the planet in a completely glacial state, with a 1\,bar N$_2$ atmosphere, and active carbonate-silicate cycle, \revvv{14\%} present-day volcanic outgassing rates, and a northern ocean land-ocean distribution. 

The \revv{schematic} trajectory illustrated in Figure \ref{fig:tempseries} provides a plausible narrative for explaining the formation of the valley networks ($\sim$4.1-3.5\,Ga) and fan/delta features ($\sim$3.3-3.0\,Ga) as episodic limit cycling that occurred in between longer periods of stable cold climates. The timing of these events is consistent with the interpretation of geologic evidence on Mars by \citet{kite2019persistence} and others that suggest fluvial features would have required globally-distributed and intense runoff, although such events could be intermittent. \rev{The abrupt temperature spikes in} the example trajectory shown in Figure \ref{fig:tempseries} \rev{are a series of} limit cycling events, \rev{which} persist for longer than the hundreds of years that would be required for such features to form. \rev{The} presence of multiple limit cycling events \rev{(sharp spikes in temperature)} within the expected temporal range for these features would further contribute to their formation. Changes in the chaotic obliquity of Mars contribute to the episodic nature of these limit cycling events, which only occur when Mars is at a relatively high obliquity (c.f., Fig. \ref{fig:obliquitylimcycle}). The particular sinusoidal function used in this simulation (Eq. \ref{eq:obliquityfunction}) is a simplification of how the obliquity of Mars would evolve on such long timescales, and the use of this function results in three separate episodes of limit cycling in this trajectory, with the most recent one occurring from about 2.4-1.8\,Ga. This \revv{schematic} trajectory does not necessarily claim that such a later episode of limit cycling must have necessarily occurred on Mars; however, it is also worth entertaining the possibility that such later limit cycling contributed to the persistence of intermittent runoff events during the later history of Mars. Even if Mars did not remain continuously warm and wet during the Noachian and Hesperian periods, multiple episodes of limit cycling could have enabled sufficiently intense and frequent runoff events to cause significant fluvial alteration to the surface and enable the formation of the geologic features observed today.

It is important that simulations performed with simple models, like those shown in Figure \ref{fig:tempseries}, are carefully interpreted and not over-extended in their application. The use of a latitudinal energy balance model in this case is computationally beneficial, as it allows for explicit calculation of the climate evolution of Mars over a multi-billion year history. Such a feat could not be performed with a more computationally sophisticated three-dimensional model. However, more complex models also offer greater physical realism, including the ability to explicitly calculate precipitation cycles and represent geographical distributions with varying latitude and longitude. The general results obtained in this study may remain robust when examined with more computationally expensive models, such as the dependence of limit cycling on obliquity and the land-ocean distribution. However, other results such as the timing of limit cycling events, the requirements for additional greenhouse gas forcing, and the duration of precipitation-driven runoff events will likely differ when approached with a three-dimensional model applicable to early Mars. Likewise, the model used in this study neglects important physical processes, such as explicit representation of cloud cover or atmospheric dynamics, which would also alter the results shown here. Although it may be computationally infeasible to use a complex three-dimensional model to re-create the full time-forward trajectory calculated here, it would be worthwhile in subsequent studies to investigate further the most significant factors that would enable Mars to enter periods of episodic limit cycling during its evolution. Such a \revv{schematic} trajectory is at least one plausible narrative that remains consistent with available martian geologic data, and further investigation by both theoretical models and Mars exploration should continue to understand the extent to which episodic, rather than persistent, runoff events could provide a better explanation for fluvial features on Mars.

\section{Conclusion}

Explaining the observation of fluvial features on Mars remains an ongoing challenge, and it is likely that multiple mechanisms contributed to their formation during the Noachian and Hesperian periods, and perhaps even into the early Amazonian. The calculations in this paper demonstrate the hypothetical viability of episodic limit cycling as an explanation for the timing of the formation of these fluvial features. The episodic limit cycling hypothesis for early Mars requires additional greenhouse gas forcing beyond CO$_2$, which could include mixtures of H$_2$, CH$_4$, and other constituents that have been suggested as possibilities. The source of additional warming could also arise from impactors, either as an alternative or a complement to additional greenhouse gases. Changes in obliquity and the land-ocean distribution over time likewise contribute to the susceptibility of early Mars to limit cycling. 
\rev{The limit cycling hypothesis for Mars could be tested to an extent through ongoing observations on Mars by improving the estimated timescales for the formation of valleys and lakes, placing better constraints on the duration and number of warming events, \revvv{constraining the volcanic outgassing rate across the history of Mars,} and understanding the extent to which such changes were representative of global conditions.}

This study is intended as a demonstration that such behavior is a feasible hypothesis to consider for the climate history of Mars, but it does not purport to be a comprehensive examination of all combinations of mechanisms that could have contributed to such a history. It even remains possible that Mars sustained standing liquid water during portions of its history but experienced limit cycles at other times. Ultimately, martian geology itself will provide the best source of constraints for these models, and further creative thinking about the climate history of Mars will undoubtedly be needed as such exploration continues.

\section*{Acknowledgments}
\revvv{Thanks to two anonymous reviewers for providing feedback that significantly improved this paper.} The author gratefully acknowledges funding from the NASA Habitable Worlds program under award 80NSSC20K0230. Any opinions, findings, and conclusions or recommendations expressed in this material are those of the author and do not necessarily reflect the views of any employer or NASA.

\bibliographystyle{elsarticle-harv} 
\bibliography{main}

\end{document}